\documentstyle[prl,aps,multicol,amssymb,amsmath,epsf]{revtex}

\def\tr{{\mathrm{Tr}}}
\def\unity{\mbox{\small 1} \!\! \mbox{1}}

\begin{document}

\title{Suitability versus fidelity for rating single-photon guns} 
\author{George M.\ Hockney, Pieter Kok, and Jonathan P.\ Dowling}
\address{Quantum Computing Technologies Group, Section 367,
  Jet Propulsion Laboratory, California Institute of Technology\\
  Mail Stop 126-347, 4800 Oak Grove Drive, Pasadena, California 91109}
\date{Version: \today}

\maketitle

\begin{abstract}
 The creation of specified quantum states is important for most, if not
 all, applications in quantum computation and communication. The
 quality of the state preparation is therefore an essential ingredient
 in any assessment of a quantum-state gun. We show that the 
 fidelity, under the standard definitions  is not sufficient to assess 
 quantum sources, and we propose
 a new measure of suitability that necessarily depends on the
 application for the source. We consider the performance of
 single-photon guns in the context of quantum key distribution (QKD)
 and linear optical quantum computation. Single-photon sources for QKD
 need radically different properties than sources for quantum computing.
 Furthermore, the suitability for single-photon guns is discussed
 explicitly in terms of experimentally accessible criteria.

 \medskip

\end{abstract}

\medskip

\begin{multicols}{2}

One of the requirements for quantum computation and communication is
the ability to faithfully produce certain input states
\cite{divincenzo00}. For quantum computers in general, this means that
we have to be able to initialize the registers in some $|0\rangle$
state. Quantum communication involves the transmission of quantum
states, and the quality of the state preparation determines in part
the success of the communication.

Optical implementations of quantum communication and computation such
as cryptography, teleportation, linear optical quantum computers,
interferometric quantum non-demolition measurements, and many other
applications, often rely on good single-photon sources \cite{apps}. 
There are many proposals for single-photon guns, ranging from
semiconductor quantum dots and manipulated individual
molecules to parametric down-converters \cite{spg}, but the
suitability of these sources has not yet been sufficiently
addressed. Whereas attenuated coherent states might be a good
approximation to a single-photon state for some applications, its
two-photon contribution renders it unsuitable for cryptography
\cite{brassard00}. Any measure of suitability therefore has to take
the intended purpose of the state into account. 

One possible choice for the suitability of a source would be the
fidelity $f_{AB}$ of the input state, with density
matrix $\rho_A$, with respect to the
desired state $\rho_B$: $f_{AB}\equiv\{\tr[(\sqrt{\rho_A}\, \rho_B\,
\sqrt{\rho_A})^{1/2}]\}^2$ \cite{jozsa94}. This fidelity satisfies
$0\leq f_{AB}\leq 1$, and most importantly, $f_{AB}=1$ if and only if
$\rho_A = \rho_B$. When one of the systems is in a pure state
$|\psi\rangle$ and the other system is in a mixed state $\rho$, this
measure reduces to $\tr(\rho|\psi\rangle\langle\psi|)$. However,
$f_{AB}$ is generally {\em not} a good measure for the suitability of
a state-preparation device (such as a single-photon gun) given a specific
application. The reason is that several different input states may be
equally suitable for a specific application. When two distinct states
$\rho_X$ and $\rho_Y$ are both useful states for a specific
application, a state $\rho_A$ might have a large $f_{XA}$, but a small
$f_{YA}$. Similarly, a state $\rho_B$ might have a small $f_{XB}$, but
a large $f_{YB}$. Both $\rho_A$ and $\rho_B$ are suitable for the
application, but the fidelity acknowledges only one of them.

In this paper we propose a general definition for the suitability
$S_{GT}$ of a source gun $G$ given a target application $T$. The
suitability satisfies $0 \leq S_{GT} \leq 1$, and is a good measure of
how well a candidate gun will work in a given target application. We
will first give an example where the fidelity breaks down as a
performance measure of the single-photon gun, and subsequently we
define the general suitability $S_{GT}$. We then use this formalism to
describe single-photon guns in quantum key distribution (QKD) and
demonstrate how a source highly suitable for QKD is not at all
suitable for quantum teleportation. 

To understand why conventional definitions of the fidelity are
problematic, consider the QKD protocol associated with the 1984
cryptographic scheme by Bennett and Brassard \cite{bennett84}. In this
application, Alice chooses a random string from a set of four
pair-wise orthogonal polarization states ${\mathcal{S}} = \{|H\rangle,
|V\rangle, |L\rangle, |R\rangle\}$. She then prepares and sends these
states to Bob. In order to ensure security, it is important that the
states only contain a single photon \cite{brassard00}. However, it is
{\em not} important that the state's frequency or exact timing within
a window is controlled (as long as this is not correlated with the
polarization). Pure states with some spread in time and frequency are
as suitable for key distribution as a mixed state, provided the
polarization is definite and the state contains exactly one photon. 

Suppose for simplicity that the unknown and unimportant part of the
state can be in only three (pure) states $|A\rangle$, $|B\rangle$, or
$|C\rangle$. When Alice wishes to send the state $|\psi\rangle \in
{\mathcal{S}}$ to Bob, it does not matter whether the state she
produces is actually $|\psi\rangle\otimes|A\rangle$,
$|\psi\rangle\otimes|B\rangle$, $|\psi\rangle\otimes|C\rangle$ or a
combination of these. Although one could require Alice to create the state
$|\psi\rangle \otimes |A\rangle$, this is overly restrictive. In a
real system this would be equivalent to requiring the input gun to
have no time jitter or any other irrelevant mixing. It only matters
that $|\psi\rangle$ is the correct polarization and has only one
photon. When we define Alice's target state as $\rho_T = \frac{1}{3}
|\psi,A\rangle\langle\psi,A| + \frac{1}{3}|\psi,B\rangle\langle\psi,B|
+ \frac{1}{3} |\psi,C\rangle\langle\psi,C|$, $\tr(\rho_G\rho_T)$ is
always smaller than 1/3 and yields different values for equally useful
input states. Using the fidelity instead, namely, $f_{AB}=\{\tr[(\sqrt{\rho_A}
\, \rho_B\, \sqrt{\rho_A})^{1/2}]\}^2$, also fails, since fidelity is 
only a measure of how much two states are alike. $f_{AB}$ is unity only
if the input state is exactly $\rho_T$, but any of several other
states are equally suitable. 

The solution is to define a procedure for writing a target state
$\rho_T$, a gun state $\rho_G$, and an expression $S(\rho_T, \rho_G)$
that has the desired propery of being close to one if and only if the
gun is suitable for the target. We define $\rho_G$, the gun state,
simply as the state that produced by the gun.  The target state $\rho_T$ 
is a bit subtler because it is not a physical state in the system.  The
procedure for defining $\rho_T$ is to identify all states which would
be useable gun states, find a set of states spanning that space,
and then define $\rho_T$ as a complete mixture of those states.
That is, given a complete set of $N$ suitable states $|\phi_i\rangle$, the
target density matrix would be $\rho_T \equiv N^{-1}\sum^N_i
|\phi_i\rangle\langle\phi_i|$. In the QKD example, 
this leads to Alice's target
state $\rho_T = \frac{1}{3} |\psi,A\rangle\langle\psi,A| +
\frac{1}{3}|\psi,B\rangle\langle\psi,B| + \frac{1}{3}
|\psi,C\rangle\langle\psi,C|$.  
To define the suitability of a quantum source, let $F_{AB} \equiv
\tr(\rho_A \rho_B)$ for any two normalized density matrices $\rho_A$
and $\rho_B$. Given two mixtures we have $F_{AB} \leq F_{AA}$. Note
that $F$ is not a proper definition of fidelity, since
$F_{AA}\neq 1$ if $\rho_A$ is not a pure state.

We propose the following definition for the suitability $S$ of a gun
that purports to create a particular quantum state:
\begin{equation}
  S_{GT} \equiv \frac{F_{GT}}{F_{TT}}\; ,
\end{equation}
where $\rho_G$ is the output state of the gun and $\rho_T$ is a
mixture of all possible target states associated with a particular
application. If there is one and only one pure target state, then $F_{TT}=1$
and $S_{GT}=f_{GT}$. For any input
state that completely overlaps the requirement, this definition of $S$ 
yields unity. This may be explicitly
worked out for the three states mentioned in the QKD example above, yielding
$F_{TT} = \frac{1}{3}$ and $S_{GT}=1$, as advertised.  

Since $\rho_T$ is a complete mixture on a subspace, the triangle
identity yields
\begin{equation}
 S_{GT} = \frac{F_{GT}}{F_{TT}} \leq \frac{F_{GG}}{F_{TT}}\; .  
\end{equation}
Since $F_{GG}$ is determined only by the gun, this leads to an
important measure of the quality of the gun that only depends on the
output state of the gun. If there is only one suitable target state
(i.e., $\rho_T$ is pure), then we have $F_{TT} = 1$ and $S_{GT}$
reduces to $F_{GT}$. This must be less than $F_{GG}$, and $F_{GG}$
therefore directly limits the suitability of a candidate gun for this
class of targets. 

So far, the formalism has been completely general, and we will now
consider the important special case of single-photon guns. Naively, an
ideal single-photon gun is a device that emits one and only one photon
with a particular frequency in a given spatial mode when triggered to
do so. Such a device can not exist. According to Heisenberg's
uncertainty relation in energy and time, it is not possible to fix
both the frequency and the triggering time with infinite
precision. This means
that we have to admit a continuum of frequency 
modes if we keep the photon number fixed in a given time window. 
The difficulties involving
ideal single-photon guns are closely related to the fact that single
photons do not have a well-defined wave function \cite{dirac28}. We
therefore have to be careful when we define the single-photon states
that are produced by the guns. In particular, the desired states
differ from application to application. 

For single-photon guns, we can modify $F_{AB}$ such
that we obtain $F^{(1)}_{AB} \equiv \tr (\rho_A |1\rangle\langle1|
\rho_B)$, where $|1\rangle\langle1| \equiv \int d\vec{k}\, d\omega\,
\hat{a}^\dag_{\vec{k}} (\omega) |0\rangle\langle0| \hat{a}_{\vec{k}}
(\omega)$. Here, $\hat{a}^{\dagger}_{\vec{k}}(\omega)$ and
$\hat{a}_{\vec{k}}(\omega)$ are the creation and annihilation
operators satisfying $[\hat{a}_{\vec{k}}(\omega),
\hat{a}^{\dagger}_{\vec{k}'}(\omega')] = \delta(\vec{k}-\vec{k}')\,
\delta(\omega-\omega')$. Obviously, we have $F_{AB}^{(1)}\leq
F_{AB}$. Inserting $|1\rangle\langle 1|$ does not affect 
$F_{GT}$ or $F_{TT}$, since $\rho_T$ is
already confined to the $|1\rangle\langle 1|$ subspace. However,
$F_{GG}^{(1)}$ gives a better measure for the performance of
single-photon guns than $F_{GG}$, since $F^{(1)}_{GG} \leq F_{GG}$,
and therefore $S_{GT} \leq F^{(1)}_{GG}/F_{TT}$.

\begin{figure}[h]
 \begin{center}
     \epsfxsize=8in
     \epsfbox[-150 20 1000 210]{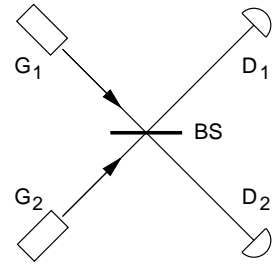}
 \end{center}
  \caption{The Hong-Ou-Mandel test: $G_1$ and $G_2$ are the
     single-photon guns and $D_1$ and $D_2$ are the
     photodetectors. When the output modes of the guns are mixed in
     the beam splitter (BS), there should be no detector coincidences
     in $D_1$ and $D_2$.} 
  \label{fig:hom}
\end{figure}

A good approximation of $F^{(1)}_{GG}$ can be measured using a
Hong-Ou-Mandel (HOM) interferometer (see Fig.\ \ref{fig:hom}). In this
measurement, the input of two identical single-photon guns $G_1$ and
$G_2$ is directed toward a beam splitter and then detected by two
ideal photodetectors $D_1$ and $D_2$. Assuming the input state from each
gun was a single photon in the same state, the output state after the
beam splitter is $|2,0\rangle - |0,2\rangle$. However, if the input
state is a mixed state or if the two guns are not identical, there
will be contamination of $|1,1\rangle$ states in the output.

The ability to trigger the gun reproducibly is an important component
of this test. If it is not possible to obtain a high $F^{(1)}_{GG}$,
then the suitability of the candidate gun for any application
requiring a pure state will be low. If the application only requires a 
mixed state (in, for example, the case of quantum key distribution
above), then the gun may still be suitable. This is a clear example
where two different applications yield different suitabilities for the
same single-photon gun. 

It is instructive to apply the suitability formalism to the HOM
experiment itself, because that demonstrates how to apply the concept of
suitability when there are two inputs. Let the input state on the beam
splitter be $\rho_{\rm in}$ and the output state $\rho_{\rm out}$. The
objective of the experiment is to obtain a high visibility in the
interference, i.e., to maximize $v = 1-\langle 1,1|\rho_{\rm out}
|1,1\rangle$, where $|1,1\rangle$ denotes the state that yields a
detector coincidence in $D_1$ and $D_2$. 

We now want to find the target state $\rho_T$ that is to be used in
determining the suitability of the single-photon guns for the HOM
experiment. On the detector side of the beam splitter, a suitable
state is either two photons in one detector or two photons in the
other.  If we label the two detectors $a$ and $b$ this is simply
\begin{eqnarray}
 \rho_T &=& \int\left\{ \left[ 
 \hat{a}_{\vec{k}}^{\dagger}(\omega) \hat{a}_{\vec{k'}}^{\dagger}
 (\omega')\, |0\rangle\langle 0|\, \hat{a}_{\vec{k}}(\omega)
 \hat{a}_{\vec{k'}}(\omega') \right] \right.+ \cr
 && \qquad \left.\left[ \hat{b}_{\vec{k}}^{\dagger}(\omega)
 \hat{b}_{\vec{k'}}^{\dagger} (\omega')\, |0\rangle\langle 0|\,
 \hat{b}_{\vec{k}}^{\dagger}(\omega) \hat{b}_{\vec{k'}}^{\dagger}
 (\omega')\right]\right\} dX \; ,   
\end{eqnarray}
where $\hat{a}_{\vec{k}}$ and $\hat{b}_{\vec{k}}$ are the annihilation
operators associated with the detectors, and $dX \equiv
h(\vec{k},\omega)\, h(\vec{k'},\omega')\,d\vec{k}\,d\omega\,
d\vec{k'}\,d\omega'$. Here $h(\vec{k},\omega)$ limits the range of $k$
and $\omega$ accepted by the beam splitter.
This is the target on the detector side of the beam splitter. We would 
prefer to have the target on the source side of the beam splitter, and
because the beam splitter is presumed to be lossless this won't change
any suitability (because of the asymmetry between the source and target,
any losses must be computed as part of the source).  If the inputs
are $c$ and $d$ this becomes 
\begin{eqnarray}
 \rho_T &=& \int \left\{
 \left[ \hat{c}_{\vec{k}}^{\dagger}
 (\omega) \hat{d}_{\vec{k'}}^{\dagger} (\omega')\, |0\rangle\langle 0|\,
 \hat{c}_{\vec{k}}(\omega) \hat{d}_{\vec{k'}}(\omega') \right]\right. + \cr 
 && \quad \frac{1}{2} \left[\hat{c}_{\vec{k}}^{\dagger} 
   (\omega) \hat{d}_{\vec{k'}}^{\dagger} (\omega') +
   \hat{d}_{\vec{k}}^{\dagger} 
   (\omega) \hat{c}_{\vec{k'}}^{\dagger} (\omega'))\right] |0\rangle \cr
 && \quad \left. \times\langle 0|\left[ (\hat{c}_{\vec{k}}
   (\omega) \hat{d}_{\vec{k'}} (\omega') +
   \hat{d}_{\vec{k}}
   (\omega) \hat{c}_{\vec{k'}} (\omega')) \right]\right\} dX\; 
\end{eqnarray}

This includes both the expected $|1,1\rangle$ inputs as well as
$|2,0\rangle + |0,2\rangle$, an eigenvector of the beam splitter.  For
the purpose of testing two independent single-photon guns it is easy
to believe the entangled state will not be generated, and so the HOM
test shows how well a pure $|1,1\rangle$ state was generated .

The input state $\rho_G = \rho_{G1}\otimes\rho_{G2}$ is the physical state
produced when the experimenter pushes the button telling {\em both} guns
to fire.  The suitability $S_{GT} = F_{GT}/F_{TT}$ is now close to unity 
if have both guns individually
generate the same pure state. Other suitable systems may involve 
either entangled or classical correlations between the two guns, such
as in parametric down-conversion. All of the required information is
always in the state $\rho_G$ of the two-gun output.   If the guns
are independent it is important that each gun's output be a pure state
and not a mixed state traced over entangled partners left in the gun.

\begin{figure}[h]
 \begin{center}
     \epsfxsize=8in
     \epsfbox[-30 20 1000 270]{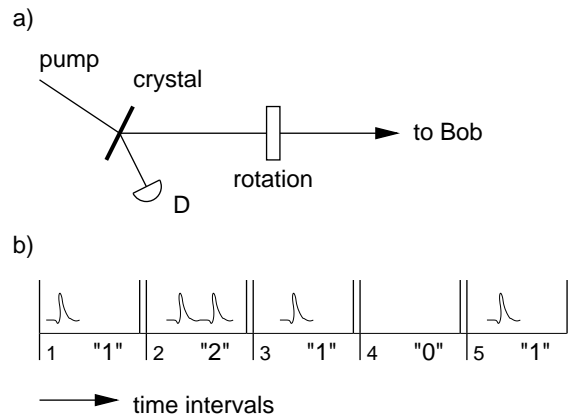}
 \end{center}
  \caption{Spontaneous parametric down-conversion in continuous-wave
     operation. a) The pump beam created photon pairs in two spatial
     modes, one of which is continuously monitored. b) The photon
     counts in the time windows. Only when one photon per time
     interval is counted by the detector $D$, we deem the creation of
     a ``single-photon state'' a success.}  
  \label{fig:spdc}
\end{figure}

As an example, the formalism developed above is used to design a QKD
system using spontaneous parametric down-conversion in continuous wave
operation (see Fig.\ \ref{fig:spdc}a). A (nearly perfect)
photodetector in the idler mode counts the number of photons per time
interval. When the outcome is ``1", the same time interval in the
signal mode then also contains exactly one photon in a known 
polarization.  Alice will rotate this photon into one of the states
$|\psi\rangle\in{\mathcal{S}}$ and send it to Bob (see Fig.\
\ref{fig:spdc}b). The state sent to Bob will have a low $F^{(1)}_{GG}$
but high $S_{GT}$. The source will have high immunity to eavesdropping
attacks, which means that it has a low suitability $S_{GE}$ for the
eavesdropper (where the target state $\rho_E$ is constructed for
maximum possible information gain by Eve). The state $\rho_G$ is now a
highly mixed state that contains only one photon (if $D$ is nearly
perfect) in the state $|\psi\rangle,$ but crossed with a large
subspace ${\mathcal{R}} = \{|A\rangle, |B\rangle, |C\rangle \dots\}$, 
representing the unknown time of emission and uncertainty in
frequency. This state has a small $F^{(1)}_{GG},$ as may be
demonstrated experimentally in a Hong-Ou-Mandel experiment. The target
state $\rho_T$ required for QKD is the mixed state
$|\psi\rangle\langle\psi| \otimes \rho_{\mathcal{R}}$, where
$\rho_{\mathcal{R}}=\unity/d$ is the identity operator on the subspace
${\mathcal{R}}$ with dimension $d$. It is clear that $F^{(1)}_{GG}
\approx 1/d$ but so are $F_{TT}$ and $F_{GT}$. Therefore, this gun has
$S_{GT} \approx 1$ and is suitable for QKD.  

Let us now consider the suitability of the single-photon gun for an
eavesdropper Eve. Since Eve does not know the polarization chosen by
Alice, she must assign her own density matrix $\sigma_G$ (i.e., her
knowledge about the protocol) to the system she intercepts. For
clarity, $\rho$ is the state according to Alice, whereas $\sigma$ is the
state according to Eve. For example, Alice creates a state with
definite polarization (say $\rho=|\psi\rangle\langle\psi|$), whereas
Eve must describe it as a maximally mixed state in the polarization
($\sigma=\unity/2$). Eve can gain information about the key when
$\sigma$ is no longer of this form. 

The target density operator $\sigma_E$ is then defined such that it
allows Eve to obtain maximal information about the string of qubits
that Alice sent to Bob. The quantity $S_{GE}=S(\sigma_G,\sigma_E)$ is
the suitability of $\sigma_G$ for eavesdropping. Perfect key
distribution is defined as $S_{GE} = 0.$ 
Notice that $S_{GE}$ is not simply related to $S_{GT}$. For example,
the source may sometimes fail to emit anything, which reduces both
$S_{GT}$ and $S_{GE}$. In this case a small $S_{GT}$ does not
necessarily mean  that eavesdropping is possible. Suppose that in a
practical system  based on this scheme the main imperfection is the detector
inefficiency (and not the possible correlations between $\mathcal{S}$
and $\mathcal{R}$). Then it is easy to calculate how a given
efficiency of $D$ affects both $S_{GT}$ and $S_{GE}$. Due to an
imperfect detector, the state $\rho_G$ acquires a contamination of
two-photon states at a fraction $\epsilon$, which would make $S_{GT} =
1 - \epsilon$ and $S_{GE} = \epsilon$. This analysis would be
completely different for other applications.  Suppose that a pure
target state were required for some protocol, for example in quantum
teleportation. As may be seen in the Hong-Ou-Mandel experiment, the 
continuous wave single-photon gun based on
down-conversion has an unmeasureably small $F^{(1)}_{GG}$. Therefore,
such a gun would be useless for any protocol requiring a 
pure target state because 
$F_{TT}\simeq 1$ and $S_{GT} \leq  F^{(1)}_{GG}.$ The
suitability of a single-photon gun therefore depends critically on the
application.


In conclusion, we have shown that the performance of a quantum source
must be evaluated in the context of a specific application. The suitability of
a source cannot be defined as the fidelity of the output state with a
target state. The reason is that there might be multiple distinct
target states all equally suited for the application. We defined the
suitability $S_{GT} = \tr(\rho_G\rho_T)/Tr(\rho_T\rho_T)$ instead,
where $\rho_G$ is the output state of the gun and $\rho_T$ is an equal
mixture of all suitable target states. In the case where $\rho_T$ is
pure the suitability reduces to the standard fidelity.

We explicitly investigated the case where the quantum source is a
single-photon gun. It was shown that for QKD, a system using a
continuously pumped SPDC crystal can provide the single-photon states
needed to prevent  eavesdropping even though this system would not be
useful for other applications where $F_{TT}$ is close to unity. It is important
to carefully understand whether a given application using
single-photon quantum optics requires a pure state or can be run with
any of a number of states.  The requirements on the photon source are
very different in these two cases. For many applications, an analysis
similar to the one done for the Hong-Ou-Mandel visibility requirement
must be done. This is because the apparatus uses several input photons
in different places, and $G$ is the combined state of {\em all} of
them. In these applications it is tempting to require that all the
guns emit maximally overlapping pure states. However, this is not
necessary if there is a correlation between the inputs. 

This work was carried out at the Jet Propulsion Laboratory, California
Institute of Technology, under a contract with the National Aeronautics 
and Space Administration. The authors wish to thank Deborah Jackson
for many useful discussions. P.K.\ acknowledges the United States
National Research Council. Support was received from the Office of
Naval Research, the Advanced Research and Development Activity, the National
Security Agency, and the Defense Advanced Research Projects Agency.

%
%
%
%
%
%
%

\end{multicols}
\end{document}